\documentclass[12pt]{article}
\usepackage{amsmath}
\usepackage{graphicx}
\usepackage{color}
\usepackage{orcidlink}
\usepackage{hyperref}
\hypersetup{colorlinks=true, linkcolor=blue, citecolor=blue, urlcolor=blue}
\usepackage{caption}
\usepackage{subcaption}
\RequirePackage[numbers,sort&compress]{natbib}
\begin{document}
\baselineskip=18pt

\begin{center}
{\large{\bf Rainbow gravity effects on relativistic quantum oscillator field in a topological defect cosmological space-time }}
\end{center}

\vspace{0.5cm}

\begin{center}
{\bf Faizuddin Ahmed\orcidlink{0000-0003-2196-9622}}\footnote{\textbf{faizuddinahmed15@gmail.com}}\\
\vspace{0.15cm}
{\it Department of Physics, University of Science \& Technology Meghalaya, Ri-Bhoi, Meghalaya, 793101, India}\\
\vspace{0.3cm}
{\bf Abdelmalek Bouzenada\orcidlink{0000-0002-3363-980X}}\footnote{\textbf{abdelmalekbouzenada@gmail.com (Corresponding author)}}\\
\vspace{0.15cm}
{\it Laboratory of Theoretical and Applied Physics, Echahid Cheikh Larbi Tebessi University 12001, Algeria} 
\par\end{center}

\vspace{0.6cm}

\begin{abstract}
In this paper, we investigate the quantum dynamics of scalar and oscillator fields in a topological defect space-time background under the influence of rainbow gravity's. The rainbow gravity's are introduced into the considered cosmological space-time geometry by replacing the temporal part $dt \to \frac{dt}{\mathcal{F}(\chi)}$ and the spatial part $dx^i \to \frac{dx^i}{\mathcal{H} (\chi)}$, where $\mathcal{F}, \mathcal{H}$ are the rainbow functions and $0 \leq \chi=|E|/E_p <1$ is the dimensionless parameter. We derived the radial equation of the Klein-Gordon equation and its oscillator equation under rainbow gravity's in topological space-time. To obtain eigenvalue of the quantum systems under investigations, we set the rainbow functions $\mathcal{F}(\chi)=1$ and $\mathcal{H}(\chi)=\sqrt{1-\beta\,\chi^p}$, where $p=1,2$. We solve the radial equations through special functions using these rainbow functions and analyze the results. In fact, it is shown that the presence of cosmological constant, the topological defect parameter $\alpha$, and the rainbow parameter $\beta$ modified the energy spectrum of scalar and oscillator fields in comparison to the results obtained in flat space. 
\end{abstract}

\vspace{0.3cm}


\vspace{0.3cm}


\section{Introduction}
\label{intro}

Rainbow gravity represents a semi-classical exploration of high-energy phenomena in quantum gravity. It achieves this by introducing higher-order terms in the energy-momentum dispersion relation through the use of rainbow functions. These functions, linked to the ratio between the energy of a test particle (such as a boson or fermion) and the Planck energy, lead to a breakdown of Lorentz symmetry at this particular energy scale \cite{M1,M2,M3}. The rainbow functions play a crucial role in governing deviations from the conventional relativistic, Minkowskian expressions. Consequently, to establish this theoretical framework, the spacetime metric must also exhibit energy dependence, with this dependence intensifying as the energy of the probing particle approaches the Planck scale.

The concept known as rainbow gravity, which is essentially an effective theory of quantum gravity, can be viewed as an endeavor to embody the generalized curvature of spacetime in the context of Doubly Special Relativity \cite{M1, M2, M4, M5}. In the realm of Doubly Special Relativity, modifications in dispersion relations are accompanied by alterations in Lorentz symmetries to establish an invariant energy (length) scale, alongside the conventional invariant velocity of light (especially at lower energies). As such, rainbow gravity is grounded on the foundational principle that these two quantities remain invariant, with the fixed energy scale being identified as the Planck energy \cite{M1, M2, M3, M4, M5}.

Rainbow gravity holds significant importance as it finds applications across various physical systems, bringing modifications to classical theories. An illustrative example is its integration with Friedmann-Robertson-Walker (FRW) cosmologies, explored in \cite{M6}, where the potential resolution of the Big Bang singularity is investigated. Furthermore, \cite{M7} delves into the exploration of the modified Starobinsky model and the inflationary solution to motion equations using rainbow gravity. This investigation extends to the calculation of crucial parameters such as the spectral index of curvature perturbation and the tensor-to-scalar ratio. Expanding its scope, rainbow gravity has also been employed to study phenomena like the deflection of light, photon time delay, gravitational red-shift, and the weak equivalence principle, as detailed in \cite{M8}.

Recent investigations into the behavior of relativistic Klein-Gordon (KG) and Dirac particles have delved into diverse spacetimes within the framework of rainbow gravity. Noteworthy examples include studies conducted in a cosmic string spacetime background, where Bezerra et al. \cite{M9} examined Landau levels through Schrödinger and KG equations. Additionally, Bakke and Mota \cite{M10, M11} explored Dirac oscillators, Sogut et al. \cite{M12} investigated the quantum dynamics of photons, and Kangal et al. \cite{M13} studied KG particles in a topologically trivial Gödel-type spacetime under rainbow gravity. Mustafa explored KG-Coulomb particles in a cosmic string rainbow gravity spacetime \cite{M14}, and massless KG-oscillators in Som-Raychaudhuri cosmic string spacetime within finely tuned rainbow gravity \cite{M15}. These studies aim to unravel the dynamics of both relativistic and non-relativistic particles in curved spacetime, providing insights into the intricate interplay between quantum mechanics and general relativity. The pursuit of exact or quasi-exact solutions for such systems illuminates the captivating influence of curved spacetime on the spectroscopic structure of relativistic particles, with implications in cosmology, the geometrical theory of topological defects, and the study of black holes and wormholes, among other areas.

In gravity’s rainbow, the metric describing the geometry of spacetime depends on the energy of the test particle used to probe the structure of that space-time. So, the geometry of space-time is represented by a family of energy dependent metrics forming a rainbow of metrics. In gravity’s rainbow, the energy–momentum dispersion relation is modified
by energy dependent rainbow functions, $\mathcal{F}(\chi)$ and $\mathcal{H}(\chi)$, where $ 0 \leq \chi(=E/E_p)<1$, such that \cite{GAC}
\begin{equation}
    E^2\,\mathcal{F}^2(\chi)-p^2\,\mathcal{H}^2(\chi)=m^2\,.\label{a1}
\end{equation}
As it is required that the usual energy-momentum dispersion relation is recovered in the infrared (IR) limit, these rainbow functions are required to satisfy
\begin{equation}
    \lim_{\chi \to 0}  \mathcal{F}(\chi) \to 1, \quad\quad\quad \lim_{\chi \to 0}  \mathcal{H}(\chi) \to 1.\label{a2}
\end{equation}
The energy dependent metric in gravity’s rainbow can be written as \cite{JM}
\begin{equation}
    g^{\mu\nu} (\chi)=e^{\mu}_a (\chi)\,e^{\nu}_b (\chi)\,\eta^{ab},\label{a3}
\end{equation}
where $\eta^{\mu}_a$ is the orthonormal tetrad fields, $\eta^{ab}$ is the Minkowski metric, and $E$ is the energy of the probing particle. The rainbow functions are defined using this energy $E$ that cannot exceed the Planck energy $E_p$.

In the literature of gravity’s rainbow, many proposals exist for the rainbow functions $\mathcal{F}(\chi)$ and $\mathcal{H}(\chi)$ \cite{RG1, RG2, RG3, RG4, RG5, RG6, RG7, RG8, RG9, RG10, RG11, RG12, RG13, RG14, RG15, RG16, RG17, RG18, RG19, RG20}. The choice of the rainbow functions is supposed to be based on phenomenological motivations. We use one of the most interesting and most studied rainbow functions that was proposed in Refs. \cite{GAC,GAC2,GAC3}
\begin{equation}
    \mathcal{F} (\chi)=1\quad\quad, \quad\quad \mathcal{H}=\sqrt{1-\beta\,\chi^p}\quad (p=1,2).\label{a8}
\end{equation}
The modified dispersion relation with these functions is compatible with some results from non-critical string theory, loop quantum gravity and $\kappa$-Minkowski non-commutative space-time. This modified dispersion relation has been used to study the dispersion of electromagnetic waves from gamma ray bursters \cite{GAC3}. It also solves the ultra high energy gamma rays paradox \cite{GAC4,GAC5}, and the paradox of the 20 TeV gamma rays from the galaxy Markarian 501 \cite{GAC4, GAC6}. In addition, this MDR provides stringent constraints on deformations of special relativity and Lorentz violations \cite{GAC7, GAC8}.

Addressing the Einstein-Maxwell equations, Bonnor formulated an exact static solution, discussed in detail for its physical implications \cite{Mel1}. Melvin later revisited this solution, leading to the currently recognized Bonnor-Melvin magnetic universe \cite{Mel2}. An axisymmetric Einstein-Maxwell solution, incorporating a varying magnetic field and a cosmological constant, was constructed in \cite{Mel3}. This electrovacuum solution was subsequently expanded upon in \cite{Mel3,Mel4}. This analysis primary focus on Bonnor-Melvin-type universe featuring a cosmological constant, discussed in detailed in Ref. \cite{MZ}. The line-element describing this magnetic universe in the chart $(t, \rho, \phi, z)$ is given by \cite{MZ}
 \begin{equation}
ds^2=-dt^2+dz^2+\frac{1}{2\,\Lambda}\,\Big(d\rho^{2}+\sin^{2} \rho\,d\phi^{2}\Big),\label{a5}
\end{equation}
where $\Lambda$ denotes the cosmological constant and the magnetic field strength is given by $B=\sqrt{2}\,\sin \rho$.

We now introduces a cosmic string into the above line-element by performing a coordinate transformation $\phi \to \alpha\,\phi$, where $\alpha$ is the cosmic string parameter  which produces an angular deficit with the ranges $0 < \alpha < 1$. Therefore, the above line-element (\ref{a5}) under this transformation becomes
\begin{equation}
ds^2=-dt^2+dz^2+\frac{1}{2\,\Lambda}\,\Big(d\rho^2+\alpha^2\,\sin^{2} \rho\,d\phi^{2}\Big),\label{a6}
\end{equation}
where now the magnetic field strength becomes $B=\sqrt{2}\,\alpha\,\sin \rho$ which is $\alpha$ times the original field strength but decreases by this amount.

Finally, the rainbow gravity's effect is introduced into this space-time (\ref{a6}) by replacing the temporal part $dt \to \frac{dt}{\mathcal{F}(\chi)}$ and the spatial part $dx^i \to \frac{dx^i}{\mathcal{H} (\chi)}$, where $\mathcal{F}, \mathcal{H}$ are the rainbow functions stated earlier. Therefore, the space-time (\ref{a6}) under rainbow gravity's can be described by the following line-element \cite{FAAB, FAAB2}
\begin{equation}
ds^2=-\frac{dt^2}{\mathcal{F}^2(\chi)}+\frac{1}{\mathcal{H}^2(\chi)}\Bigg[dz^2+\frac{1}{2\,\Lambda}\,\Big(d\rho^2+\alpha^2\,\sin^2 \rho\,d\phi^2\Big)\Bigg].\label{a7}
\end{equation}
One can evaluate the magnetic field strength for the modified topological defect cosmological space-time (\ref{a7}) and it is given by $\mathcal{B}=\frac{\alpha}{\sqrt{2}\,\mathcal{H}(\chi)}\,\sin \rho $ which depends on the rainbow function. For $\chi \to 0$, we will get back the topological defect cosmological space-time (\ref{a6}).

The exploration of rainbow gravity's implications in quantum mechanical problems has attracted much attention in current times. Various studies have delved into the effects of rainbow gravity in different quantum systems, including the Dirac oscillator in cosmic string space-time \cite{aa13}, scalar fields in a wormhole background with cosmic strings \cite{AG}, quantum motions of scalar particles \cite{EE}, and the behavior of spin-1/2 particles in a topologically trivial G\"{o}del-type space-time \cite{EE2}. Additionally, investigations have extended to the motions of photons in cosmic string space-time \cite{EE3}, and the generalized Duffin-Kemmer-Petiau equation with non-minimal coupling in cosmic string space-time \cite{EE4} and some others reported in Refs. \cite{QQ1,QQ3}.

Our motivation is to study the relativistic quantum motions of oscillator field in a topological defect cosmological space-time under the influence of rainbow gravity's. It is well-known that the presence of rainbow gravity's in a space-time alter its geometrical characteristics, and hence, behaviours of quantum particles in this geometry background would definitely changes. We study this quantum mechanical problems by choosing two pairs of rainbow functions by considering $p=1,2$ in Eq. (\ref{a8}). In fact, we show that the solutions of the relativistic wave equation via the Klein-Gordon oscillator are influenced by the cosmological constant $\Lambda$, the topological defect parameter $\alpha$ and the rainbow parameter $\beta$. Throughout this paper, we use the system of units, where $\hbar=c=G=1$.

\section{Quantum oscillator field in topological defect cosmological space-time }

We study dynamics of quantum oscillator field described by the Klein-Gordon oscillator under the environment of rainbow gravity's is investigated in curved space-time produced by a magnetic field given by the metric (\ref{a7}). 

The covariant ($g_{\mu\nu}$) and contravariant form ($g_{\mu\nu}$) of the metric tensor for the space-time (\ref{a7}) are given by
\begin{eqnarray}
&&g_{\mu\nu}=\mbox{diag}\Big(-\frac{1}{\mathcal{F}^2(\chi)}, \frac{1}{2\,\Lambda\,\mathcal{H}^2(\chi)}, \frac{\alpha^2\,\sin^2 \rho}{2\,\Lambda\,\mathcal{H}^2(\chi)}, \frac{1}{\mathcal{H}^2(\chi)} \Big),\nonumber\\
&&g^{\mu\nu}=\mbox{diag}\Big(-\mathcal{F}^2(\chi),  2\,\Lambda\,\mathcal{H}^2(\chi), \frac{2\,\Lambda\,\mathcal{H}^2(\chi)}{\alpha^2\,\sin^2 \rho}, \mathcal{H}^2(\chi)  \Big). \label{b2}
\end{eqnarray}
The determinant of the metric tensor for the space-time (\ref{a7}) is given by 
\begin{equation}
    det\,(g_{\mu\nu})=g=-\frac{\alpha^2\,\,\sin^2 \rho}{4\,\Lambda^2\,\mathcal{H}^6(\chi)\,\mathcal{F}^2(\chi)}\,. \label{b3}
\end{equation}

The relativistic quantum motions of scalar particles is described by the Klein-Gordon equation \cite{FAAB, FAAB2, WG}
\begin{eqnarray}
    \Bigg[-\frac{1}{\sqrt{-g}}\,\partial_{\mu}\,\Big(\sqrt{-g}\,g^{\mu\nu}\,\partial_{\nu}\Big)+M^2\Bigg]\,\Psi=0,\label{b1}
\end{eqnarray}
where $M$ is the rest mass of the particles, $g$ is the determinant of the metric tensor $g_{\mu\nu}$ with its inverse $g^{\mu\nu}$.

The Klein-Gordon oscillator equation is studied by substituting the momentum four-vector $\partial_{\mu} \to (\partial_{\mu}+M\,\omega\,X_{\mu})$, with $X_{\mu}=(0, r, 0, 0)$ and $\omega$ represents the oscillator frequency. Extensive exploration investigating the dynamics of quantum oscillator field in various space-time backgrounds including G\"{o}del and G\"{o}del-type space-times have been done. Moreover, investigations have been conducted in global monopoles, cosmic string space-times (both standard and spinning), as well as in topologically trivial and non-trivial geometries (see, Refs. \cite{JJ1, JJ2, JJ3, JJ4, JJ5, JJ6, JJ7, JJ8, ss10, JJ10, JJ11, JJ12,JJ13}). 

Therefore, the relativistic wave equation describing the quantum oscillator field is given by
\begin{eqnarray}
    \Big[\frac{1}{\sqrt{-g}}\,(\partial_{\mu}+M\,\omega\,X_{\mu})\,(\sqrt{-g}\,g^{\mu\nu})\,(\partial_{\nu}-M\,\omega\,X_{\nu})\Big]\,\Psi=M^2\,\Psi\,.  \label{e1}
\end{eqnarray}

Expressing this wave equation (\ref{e1}) in the space-time background (\ref{a7}), we obtain the following equation
\begin{eqnarray}
&&\Bigg[-\mathcal{F}^2\,\frac{d^2}{dt^2}+2\,\Lambda\,\mathcal{H}^2\,\Bigg\{\frac{d^2}{d\rho^2}+\frac{1}{\tan \rho}\,\frac{d}{d\rho}-M\,\omega-\frac{M\,\omega\,\rho}{\tan \rho}-M^2\,\omega^2\,\rho^2+\frac{1}{\alpha^2\,\sin^2 \rho}\,\frac{d^2}{d\phi^2}\Bigg\}\nonumber\\
&&+\mathcal{H}^2\,\frac{d^2}{dz^2}-M^2\Bigg]\,\Psi=0\,.\label{e2}
\end{eqnarray}

In quantum mechanical system, the total wave function is always expressible in terms of different variables. In our case, we choose the following ansatz of the total wave function $\Psi$ in terms of different variable functions $\psi (x)$ and $\Theta(\theta)$ as follows:
\begin{equation}
    \Psi =\exp(-i\,E\,t)\,\exp(i\,\ell\,\phi)\,\exp(i\,k\,z)\,\psi(\rho), \label{b5}
\end{equation}
where $E$ is the particles energy, $\ell=0,\pm\,1,\pm\,2,....$ are the eigenvalues of the angular quantum number, and $k$ is an arbitrary constant.

Substituting this function (\ref{b5}), we obtain 
\begin{equation}
\psi'' (\rho)+\frac{1}{\tan \rho}\,\psi' (\rho)+\Bigg[\frac{(\mathcal{F}^2\,E^2-M^2)}{2\,\Lambda\,\mathcal{H}^2}-M\,\omega-\frac{M\,\omega\,\rho}{\tan \rho}-M^2\,\omega^2\,\rho^2-\frac{k^2}{2\,\Lambda}-\frac{\iota^2}{\sin^2 \rho}\Bigg]\,\psi (\rho)=0.\label{e3}
\end{equation}

We solve the above equation taking an approximation of up to the first order, that is, for small values of the radial distance $\rho$. Therefore, the radial wave equation (\ref{e3}) reduces to the following form:
\begin{equation}
\psi'' (\rho)+\frac{1}{\rho}\,\psi' (\rho)+\Big(\xi^2-M^2\,\omega^2\,\rho^2-\frac{\iota^2}{\rho^2}\Big)\,\psi (\rho)=0,\label{e4}
\end{equation}
where we have defined
\begin{equation}
\xi^2=\frac{(\mathcal{F}^2\,E^2-M^2)}{2\,\Lambda\,\mathcal{H}^2}-\frac{k^2}{2\,\Lambda}-2\,M\,\omega\,.\label{e5}
\end{equation}

Transforming the above equation (\ref{e4}) to a new variable via 
\begin{equation}
    s=M\,\omega\,\rho^2\label{e6}    
\end{equation}
 we obtain
\begin{equation}
    \psi''(s)+\frac{1}{s}\,\psi'(s)+\Bigg(\frac{\xi^2}{4\,M\,\omega}\,\frac{1}{s}-\frac{1}{4}-\frac{\iota^2/4}{s^2}\Bigg)\,\psi (s)=0.\label{e7}
\end{equation}
{\color{red}
That can be written as
\begin{equation}
    \psi''(s)+\frac{(c_1-c_2\,s)}{s\,(1-c_3\,s)}\,\psi'(s)+\frac{\left(-\eta_1\,s^2+\eta_2\,s-\eta_3\right)}{s^2\,(1-c_3\,s)^2}\,\psi (s)=0,\label{ee7}
\end{equation}
where
\begin{equation}
    \eta_1=\frac{1}{4},\quad\quad \eta_2=\frac{\xi^2}{4\,M\,\omega},\quad\quad \eta_3=\frac{\iota^2}{4}.\label{ee8}
\end{equation}
And the coefficients are
\begin{equation}
    c_1=1,\quad c_2=0=c_3.\label{ee9}
\end{equation}

The above second-order homogeneous differential equation (\ref{ee7}) can be solved using a well-known method known as the Nikiforov-Uvarov (NU) method \cite{AFN}. This method has successfully been applied in solving quantum mechanical problems in the literature (see, for examples,  \cite{FAAB,JJ1,JJ2,JJ10,JJ11,JJ13}.}

Following the procedure in \cite{AFN}, the energy eigenvalue relation is given by (please see appendix for the calculation)
\begin{equation}
    \mathcal{F}^2 (\chi)\,E^2-M^2=\mathcal{H}^2(\chi)\,\Bigg[k^2+8\,M\,\omega\,\Lambda\,\Big(n+\frac{|\ell|}{2\,\alpha}+1\Big)\Bigg]\,.\label{e12}
\end{equation}
The corresponding wave function will be
\begin{equation}
    \psi_{n, \ell} (s)=s^{\frac{|\ell|}{2\,\alpha}}\,e^{-\frac{s}{2}}\,L^{\left(\frac{|\ell|}{\alpha}\right)}_n(s),\label{ee1}
\end{equation}
where $L^{\left(\frac{|\ell|}{\alpha}\right)}_{n}(s)$ is the generalized Laguerre polynomials.

In terms of $\rho$, we obtain the radial wave function
\begin{equation}
    \psi_{n, \ell} (\rho)=\mathcal{C}\,(M\,\omega)^{\frac{|\ell|}{2\,\alpha}}\,\rho^{\frac{|\ell|}{\alpha}}\,e^{-\frac{1}{2}\,M\,\omega\,\rho^2}\,L^{\left(\frac{|\ell|}{\alpha}\right)}_n(M\,\omega\,\rho^2),\label{ee2}
\end{equation}
$\mathcal{C}$ is the normalization constant.

Below, we evaluate the energy spectrum of oscillator fields by choosing two pairs of rainbow functions considered earlier. 

\vspace{0.2cm}
\begin{center}
    {\bf Case A:} $\mathcal{F} (\chi)=1$ and $\mathcal{H} (\chi)=\sqrt{1-\beta\,\chi}$, where $\chi=\frac{E}{E_p}$
\end{center}

In this case, we choose a pair of rainbow functions given by
\begin{equation}
    \mathcal{F}=1,\quad \mathcal{H}=\sqrt{1-\beta\,\chi},\quad \chi=\frac{E}{E_p}\,.\label{e13}
\end{equation}

Substituting this pair of rainbow function into the energy eigenvalue relation (\ref{e12}), we obtain
\begin{equation}
    E^2_{n\,\ell}-M^2=\Big(1-\frac{\beta}{E_p}\,|E_{n\,\ell}|\Big)\,\Bigg[k^2+8\,M\,\omega\,\Lambda\,\Big(n+\frac{|\ell|}{2\,\alpha}+1\Big)\Bigg].\label{e14}
\end{equation}

For $|E_{n\,\ell}|=E_{n\,\ell}$, simplification of the above relation gives us an approximate energy eigenvalue for particles given by
\begin{equation}
    E^{+}_{n\,\ell}=-\Big(\frac{\beta}{E_p}\Big)\,\frac{\Theta_{n\,\ell}}{2}+\sqrt{\Big(\frac{\beta}{E_p}\Big)^2\,\frac{\Theta^2_{n\,\ell}}{4}+M^2+\Theta_{n\,\ell}}>0,\label{e15}
\end{equation}

\begin{center}
\begin{figure}
\begin{centering}
\subfloat[$\alpha=0.5,\ell=1$]{\centering{}\includegraphics[scale=0.55]{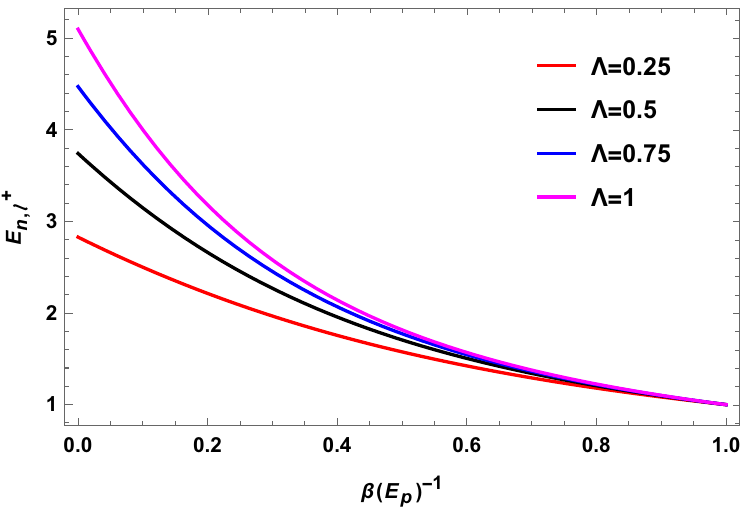}}\quad\quad\quad
\subfloat[$\varLambda=0.5,\ell=1$]{\centering{}\includegraphics[scale=0.55]{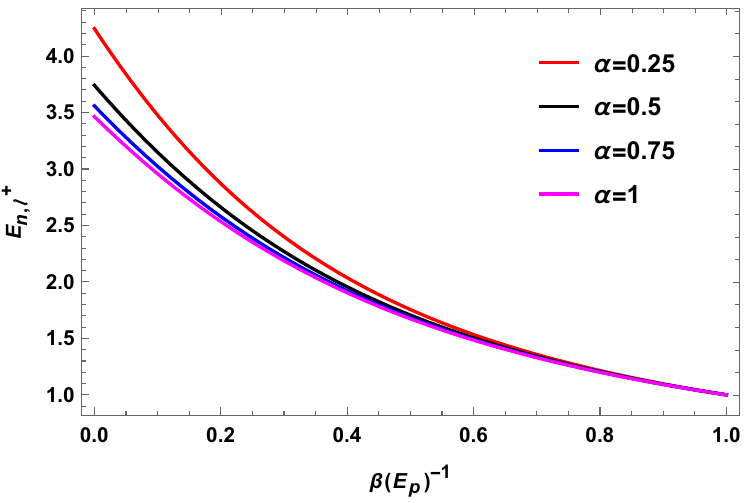}}
\par\end{centering}
\hfill\\
\begin{centering}
\subfloat[$\alpha=\Lambda=0.5$]{\centering{}\includegraphics[scale=0.55]{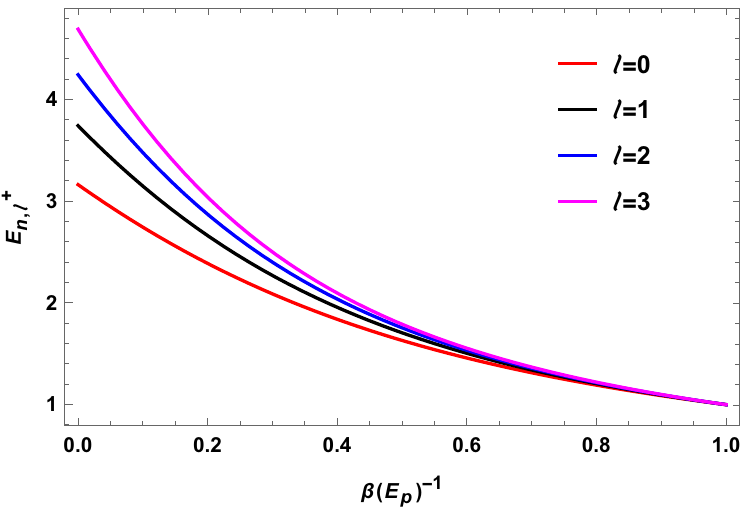}}
\par\end{centering}
\caption{Positive energy $E_{n\,\ell}^{+}$ in (\ref{e15}), where $n=\omega=k=M=1$.}
\label{Fig:1}
\end{figure}
\par\end{center}

where
\begin{equation}
    \Theta_{n\,\ell}=k^2+8\,M\,\omega\,\Lambda\,\Big(n+\frac{|\ell|}{2\,\alpha}+1\Big).\label{e16}
\end{equation}

For $|E_{n\,\ell}|=-E_{n\,\ell}$, simplification of the above relation gives us an approximate energy eigenvalue for anti-particles given by
\begin{equation}
    E^{-}_{n\,\ell}=\Big(\frac{\beta}{E_p}\Big)\,\frac{\Theta_{n\,\ell}}{2}-\sqrt{\Big(\frac{\beta}{E_p}\Big)^2\,\frac{\Theta^2_{n\,\ell}}{4}+M^2+\Theta_{n\,\ell}}=-E^{+}_{n\,\ell}<0.\label{e155}
\end{equation}

\begin{center}
\begin{figure}
\begin{centering}
\subfloat[$\alpha=0.5,\ell=1$]{\centering{}\includegraphics[scale=0.55]{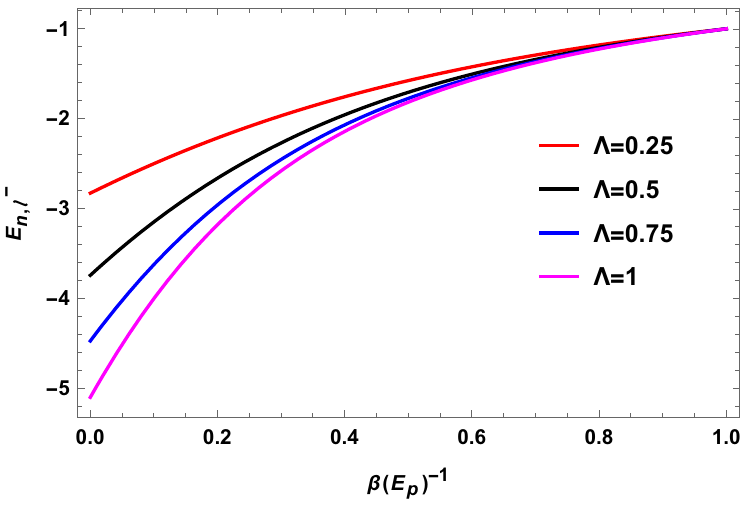}}\quad\quad\quad
\subfloat[$\varLambda=0.5,\ell=1$]{\centering{}\includegraphics[scale=0.55]{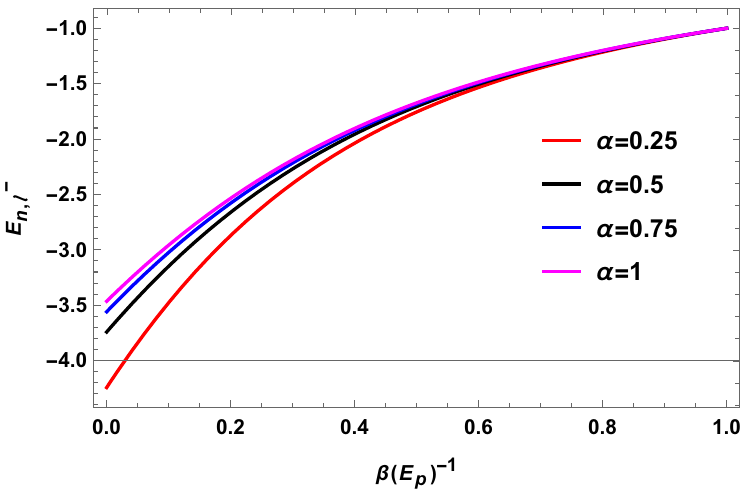}}
\par\end{centering}\
\hfill\\
\begin{centering}
\subfloat[$\alpha=\Lambda=0.5$]{\centering{}\includegraphics[scale=0.55]{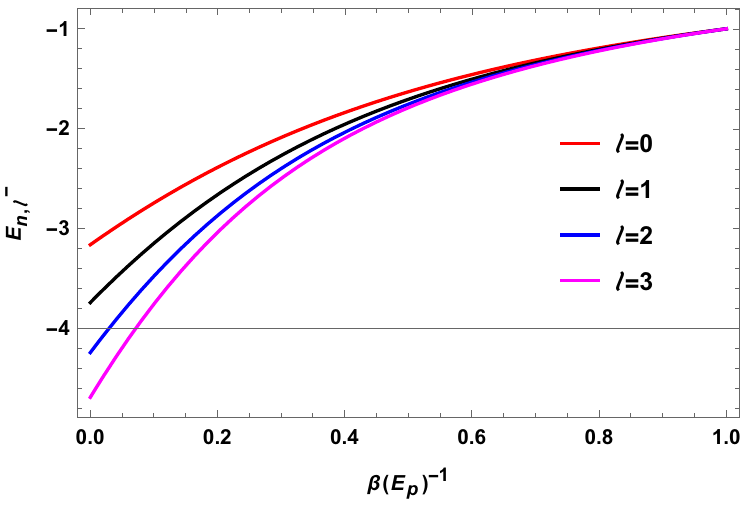}}
\par\end{centering}
\caption{The energy spectrum $E_{n\,\ell}^{-}$ of Eq. (\ref{e155}), where $n=\omega=k=M=1$.}
\label{Fig:2}
\end{figure}
\par\end{center}

Equations (\ref{e15}) and (\ref{e155}) are the relativistic approximate energy eigenvalue of quantum oscillator fields associated with the modes $\{n, \ell\}$ in the background of topological defect cosmological space-time (\ref{a7}) for the chosen rainbow function $\mathcal{F}(\chi)=1$ and $\mathcal{H}(\chi)=\sqrt{1-\beta\,\chi}$.

We have plotted the energy spectrum of quantum oscillator fields given in Eq. (\ref{e15}) in Figure \ref{Fig:1} and Eq. (\ref{e155}) in Figure \ref{Fig:2} for different values of the cosmological constant $\Lambda$, the topological parameter $\alpha$, and the angular quantum number $\ell$ and shows their behaviour with increasing values of these parameters. 

\begin{center}
\begin{figure}
\begin{centering}
\subfloat[$\alpha=0.5,\ell=1$]{\centering{}\includegraphics[scale=0.52]{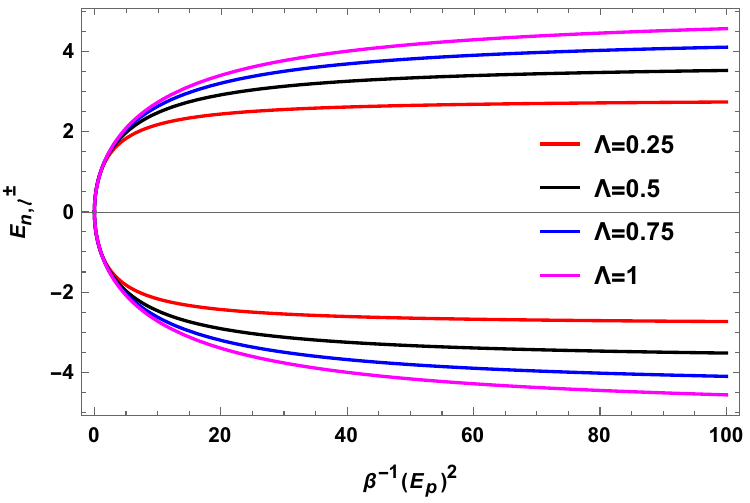}}\quad\quad\quad
\subfloat[$\varLambda=0.5,\ell=1$]{\centering{}\includegraphics[scale=0.52]{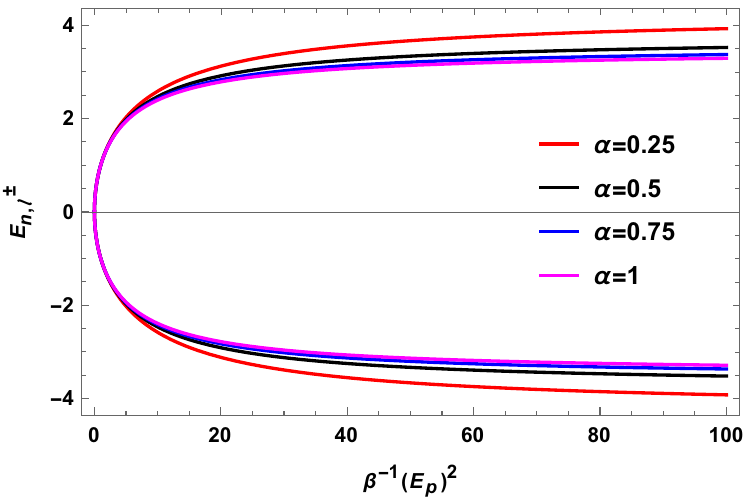}}
\par\end{centering}
\hfill\\
\begin{centering}
\subfloat[$\alpha=\Lambda=0.5$]{\centering{}\includegraphics[scale=0.52]{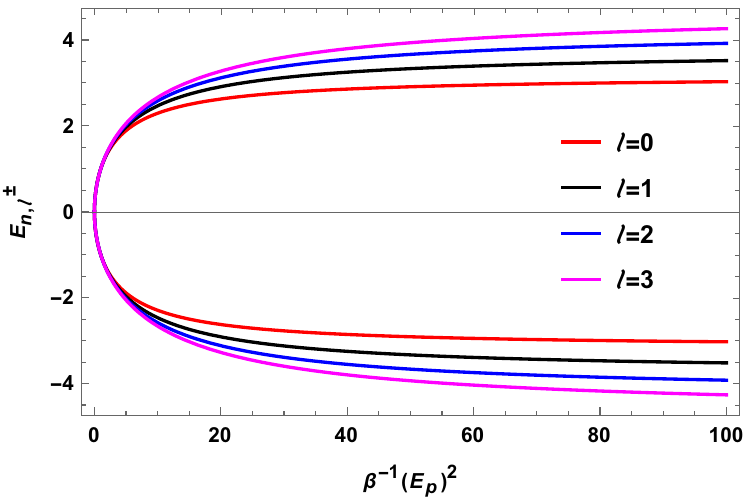}}
\par\end{centering}
\caption{The energy spectrum $E_{n\,\ell}^{\pm}$ of Eq. (\ref{e19}), where $n=\omega=k=M=1$.}
\label{Fig:3}
\end{figure}
\par\end{center}

\vspace{0.2cm}
\begin{center}
    {\bf Case B:} $\mathcal{F} (\chi)=1$ and $\mathcal{H} (\chi)=\sqrt{1-\beta\,\chi^2}$, where $\chi=\frac{E}{E_p}$
\end{center}

In this case, we choose a pair of rainbow functions given by
\begin{equation}
    \mathcal{F} (\chi)=1,\quad \mathcal{H} (\chi)=\sqrt{1-\beta\,\chi^2},\quad \chi=\frac{E}{E_p}\,.\label{e17}
\end{equation}

Substituting this pair of rainbow function into the energy eigenvalue relation (\ref{e12}), we obtain
\begin{equation}
    E^2_{n, \ell}-M^2=\Big(1-\frac{\beta}{E^{2}_p}\,E^2_{n, \ell}\Big)\,\Bigg[k^2+8\,M\,\omega\,\Lambda\,\Big(n+\frac{|\ell|}{2\,\alpha}+1\Big)\Bigg]\,.\label{e18}
\end{equation}
Simplification of the above relation gives us the compact expression of the energy eigenvalue given by
\begin{equation}
    E_{n\,\ell}^{\pm}=\pm\sqrt{\frac{M^{2}+k^{2}+8\,M\,\omega\,\Lambda\,\Big(n+\frac{|\ell|}{2\,\alpha}+1\Big)}{1+\frac{\beta}{E_{p}^{2}}\,\left[k^{2}+8\,M\,\omega\,\Lambda\,\Big(n+\frac{|\ell|}{2\,\alpha}+1\Big)\right]}}.\label{e19}
\end{equation}

Equation (\ref{e19}) is the relativistic approximate energy eigenvalue of quantum oscillator fields associated with the modes $\{n, \ell\}$ in the background of topological defect cosmological space-time (\ref{a7}) for the chosen rainbow function $\mathcal{F}(\chi)=1$ and $\mathcal{H}(\chi)=\sqrt{1-\beta\,\chi^2}$.

We have plotted the energy spectrum of the quantum oscillator fields given in Eq. (\ref{e19}) in Figure \ref{Fig:3} for different values of the cosmological constant $\Lambda$, the topological parameter $\alpha$, and the angular quantum number $\ell$ and shows their behaviour with increasing values of these parameters. 

\section{Conclusions}

We explored the relativistic quantum oscillator field within the framework of rainbow gravity in the background of cosmological space-time. By deriving the radial equation of the Klein-Gordon oscillator through a suitable wave function ansatz and utilizing special functions, particularly the Nikiforov-Uvarov (NU) method, we obtained the relativistic energy spectrum and corresponding wave function of the quantum oscillator field. This analysis was conducted by selecting two different forms of the rainbow function, which hold immense significance in cosmology. It is worth noting that the NU-method has gained considerable interest for solving quantum mechanical problems in both the relativistic and non-relativistic regimes.

We demonstrated that various parameters, such as the cosmological constant $\Lambda$, which determines the strength of the magnetic field, and the cosmic string parameter $\alpha$, influence the eigenvalue solutions of quantum systems. Moreover, the rainbow parameter $\beta$, which includes the Plancks' energy scale, modifies the energy eigenvalue of the quantum oscillator field. The study of the relativistic quantum oscillator field in curved space-time, particularly within the paradigm of rainbow gravity, broadens our understanding of how fundamental particles (spin-zero oscillator fields) behave in diverse gravitational environments. These findings contribute to the exploration of unconventional aspects of the relationship between gravity and quantum physics, offering valuable insights into the nature of quantum fields in space-time.

In future research, we plan to explore the quantum motion of spin-zero and spin-half particles in rainbow gravity backgrounds using various external potential models, such as the Yukawa potential, Poschl-Teller potential, and Morse potential. These studies will further analyze the resulting quantum systems and their implications. 

\section*{Conflict of Interest}

There is no conflict of interests regarding publication of this paper.

\section*{Funding Statement}

There is no funding agency associated with this manuscript.

\section*{Data Availability Statement}

No new data are generated or analysed during this study.

\section*{Appendix: The parametric Nikiforov-Uvarov (NU) method}\label{app}

\setcounter{equation}{0}
\renewcommand{\theequation}{A.\arabic{equation}}

The Nikiforov-Uvarov method is helpful to find exact and approximate energy eigenvalue and the wave function of the Schr\"{o}dinger-like equation and other second-order differential equations of physical interest. According to this method, the wave functions of a second-order differential equation \cite{AFN}
\begin{eqnarray}
\psi''(s)+\frac{(c_1-c_2\,s)}{s\,(1-c_3\,s)}\psi'(s)+\frac{(-\xi_1\,s^2+\xi_2\,s-\xi_3)}{s^2\,(1-c_3\,s)^2}\psi (s)=0
\label{A.1}
\end{eqnarray}
are given by 
\begin{eqnarray}
\psi (s)=s^{c_{12}}(1-c_3 s)^{-c_{12}-\frac{c_{13}}{c_3}}\,P^{(c_{10}-1,\frac{c_{11}}{c_3}-c_{10}-1)}_{n}(1-2\,c_3 s).
\label{A.2}
\end{eqnarray}
And that the energy eigenvalue equation is given by
\begin{eqnarray}
c_2\,n-(2\,n+1)\,c_5+(2\,n+1)\,(\sqrt{c_9}+c_3\,\sqrt{c_8})+n\,(n-1)\,c_3+c_7+2\,c_3\,c_8+2\,\sqrt{c_8\,c_9}=0.\quad\quad
\label{A.3}
\end{eqnarray}
The parameters $c_4,\ldots,c_{13}$ are obatined from the six parameters $c_1,\ldots,c_3$ and $\xi_1,\ldots,\xi_3$ as follows:
\begin{eqnarray}
&&c_4=\frac{1}{2}\,(1-c_1),\quad c_5=\frac{1}{2}\,(c_2-2\,c_3),\quad
c_6=c^2_{5}+\xi_1,\quad c_7=2\,c_4\,c_{5}-\xi_2,\quad c_8=c^2_{4}+\xi_3,\nonumber\\
&&c_9=c_6+c_3\,c_7+c^{2}_3\,c_8,\quad c_{10}=c_1+2\,c_4+2\,\sqrt{c_8},\quad
c_{11}=c_2-2\,c_5+2\,(\sqrt{c_9}+c_3\,\sqrt{c_8}),\nonumber\\
&&c_{12}=c_4+\sqrt{c_8},\quad c_{13}=c_5-(\sqrt{c_9}+c_3\,\sqrt{c_8}).
\label{A.4}
\end{eqnarray}

A special case where $c_3=0$, we find
\begin{equation}
\lim_{c_3\rightarrow 0} P^{(c_{10}-1,\frac{c_{11}}{c_3}-c_{10}-1)}_{n}\,(1-2\,c_3\,s)=L^{c_{10}-1}_{n} (c_{11}\,s),
\label{A.5}
\end{equation}
and 
\begin{equation}
\lim_{c_3\rightarrow 0} (1-c_3\,s)^{-c_{12}-\frac{c_{13}}{c_3}}=e^{c_{13}\,s}.
\label{A.6}
\end{equation}
Therefore the wave-function from (\ref{A.2}) becomes
\begin{equation}
\psi (s)=s^{c_{12}}\,e^{c_{13}\,s}\,L^{c_{10}-1}_{n} (c_{11}\,s),
\label{A.7}
\end{equation}
where $L^{\beta}_{n} (x)$ denotes the generalized Laguerre polynomial. 

The energy eigenvalues equation reduces to 
\begin{equation}
n\,c_2-(2\,n+1)\,c_5+(2\,n+1)\,\sqrt{c_9}+c_7+2\,\sqrt{c_8\,c_9}=0.
\label{A.8}
\end{equation}

\end{document}